\documentclass[prb,preprint,superscriptaddress,noshowpacs,noshowkeys,endfloats*]{revtex4}

\usepackage{graphicx}
\usepackage{graphics}
\usepackage{bm}
\usepackage{dcolumn}
\usepackage{amsmath}
\usepackage{bm}
\usepackage{color}

\begin{document}

\def\BiSe{Bi$_2$Se$_3$}
\def\BiSeEuSe{Bi$_2$Se$_3$/EuSe}
\def\BiSeEuS{Bi$_2$Se$_3$/EuS}
\def\Tb{$T_{\rm C,B}$}
\def\Ts{$T_{\rm C,S}$}
\def\Tn{$T_{\rm N}$}
\def\BiMnSe{(Bi$_{1-x}$Mn$_x$)$_2$Se$_3$}
\def\Tc{$T_{\rm C}$}
\def\BiTe{Bi$_2$Te$_3$}
\def\BiSe{Bi$_2$Se$_3$}
\def\SbTe{Sb$_2$Te$_3$}
\def\BiSbTe{(Bi, Sb)$_2$Te$_3$}
\def\SbVTe{(Sb$_{1-x}$V$_x$)$_2$Te$_3$}
\def\BiMnTe{(Bi$_{1-x}$Mn$_x$)$_2$Te$_3$}
\def\BiMnSe{(Bi$_{1-x}$Mn$_x$)$_2$Se$_3$}
\def\Gbar{\overline{$\Gamma$}}
\def\MnBi{Mn$_{\rm Bi}$} 
\def\muB{$\mu_{\rm B}$}

\title{Search for enhanced magnetism at the interface between \BiSe\ and EuSe}

\author{K.~Proke\v{s}}
\email{prokes@helmholtz-berlin.de}
\affiliation{Helmholtz-Zentrum Berlin f\"{u}r Materialien und Energie, Hahn-Meitner-Platz 1, 14109 Berlin, Germany}

\author{Chen Luo}
\affiliation{Helmholtz-Zentrum Berlin f\"{u}r Materialien und Energie, Albert-Einstein-Stra$\ss$e 15, 12489 Berlin, Germany}
\affiliation{Fakult\"at f\"ur Physik, Technische Universit\"at M\"unchen, James-Franck-Stra$\ss$e 1, 85748 Garching bei M\"{u}nchen, Germany}

\author{H. Ryll}
\affiliation{Helmholtz-Zentrum Berlin f\"{u}r Materialien und Energie, Albert-Einstein-Stra$\ss$e 15, 12489 Berlin, Germany}

\author{E. Schierle}
\affiliation{Helmholtz-Zentrum Berlin f\"{u}r Materialien und Energie, Albert-Einstein-Stra$\ss$e 15, 12489 Berlin, Germany}

\author{D. Marchenko}
\affiliation{Helmholtz-Zentrum Berlin f\"{u}r Materialien und Energie, Albert-Einstein-Stra$\ss$e 15, 12489 Berlin, Germany}

\author{E. Weschke}
\affiliation{Helmholtz-Zentrum Berlin f\"{u}r Materialien und Energie, Albert-Einstein-Stra$\ss$e 15, 12489 Berlin, Germany}

\author{F. Radu}
\affiliation{Helmholtz-Zentrum Berlin f\"{u}r Materialien und Energie, Albert-Einstein-Stra$\ss$e 15, 12489 Berlin, Germany}

\author{R. Abrudan}
\affiliation{Helmholtz-Zentrum Berlin f\"{u}r Materialien und Energie, Albert-Einstein-Stra$\ss$e 15, 12489 Berlin, Germany}

\author{V. V. Volobuev}
\altaffiliation[Present address:]{ International Research Centre MagTop and Institute of Physics, Polish Academy of Sciences, Aleja Lotnik\'ow 32/46, 02668 Warsaw, Poland}
\affiliation{Institute for Semiconductor and Solid State Physics, Johannes Kepler Universit\"{a}t, 4040 Linz, Austria}

\author{G. Springholz}
\affiliation{Institute for Semiconductor and Solid State Physics, Johannes Kepler Universit\"{a}t, 4040 Linz, Austria}

\author{O. Rader}
\affiliation{Helmholtz-Zentrum Berlin f\"{u}r Materialien und Energie, Albert-Einstein-Stra$\ss$e 15, 12489 Berlin, Germany}

\date{\today}

\begin{abstract}  
Enhanced magnetism has recently been reported for the topological-insulator/ferromagnet interface \BiSeEuS\ with Curie temperatures claimed to be raised above room temperature from the bulk EuS value of 16 K. Here we investigate the analogous interface \BiSeEuSe. EuSe is a low-temperature layered ferrimagnet that is particularly sensitive to external perturbations. We find that superconducting quantum interference device (SQUID) magnetometry of \BiSeEuSe\ heterostructures reveals precisely the magnetic phase diagram known from EuSe, including the ferrimagnetic phase below 5 K, without any apparent changes from the bulk behavior. Choosing a temperature of 10 K to search for magnetic enhancement, we determine an upper limit for a possible magnetic coercive field of 3 mT. Using interface sensitive x-ray absorption spectroscopy we verify the magnetic divalent configuration of the Eu at the interface without contamination by Eu$^{3+}$, and by x-ray magnetic circular dichroism (XMCD) we confirm at the interface the magnetic hysteresis obtained by SQUID. XMCD data obtained at 10 K in a magnetic field of 6 T indicate a magnetic spin moment of $m_{z,{\rm spin}} = 7$ $\mu_{B}$/Eu$^{2+}$, in good agreement with the SQUID data and the expected theoretical  moment of Eu$^{2+}$. Subsequent XMCD measurements in zero field show, however, that sizable remanent magnetization is absent at the interface for temperatures down to about 10 K. 
\end{abstract}

\maketitle

\section{Introduction}
 

At the surface or interface of ferromagnets, the magnetic order may be modified. Particularly interesting are cases in which the ferromagnetic  ordering temperature is enhanced over its bulk value. Such a behavior has been predicted by a nearest-neighbor Ising model under the condition that the magnetic coupling constant at the surface $J_{\rm S}$ exceeds that of the bulk by a certain critical value, upon  which the surface orders at a temperature \Ts\ higher than the bulk Curie temperature \Tb.\cite{Binder83}  
Such enhanced magnetic interaction may be possible with topological insulators (TI) \cite{Moore10,Hasan10} due to the presence of the topological state localized at the surface. In particular, it has been predicted for the  topological insulator \BiSe\ that this topological surface state couples magnetic moments ferromagnetically through an RKKY-type interaction.\cite{LiuQPRL09,BiswasPRB10} 
 
One candidate material has been \BiMnSe. Mean-field theory calculations examined the possible enhancement of the Curie temperature for this system.\cite{Rosenberg12}
 A representative prediction out of several models was an enhancement from \Tb$=73$ K   to \Ts$ = 103$ K.\cite{Rosenberg12} However, in experiment, the expected enhancement has not been confirmed, indicating that magnetically doped topological insulators may not be suited for the observation of enhanced magnetism.\cite{SanchezNC16,Rienks19}

Recently, two intriguing observations have been reported for such interfaces: on the one hand a giant enhancement of the spin-orbit torque\cite{Mellnik14,Fan14,Wang15,Fan16}, 
and on the other hand, a giant enhancement of \Tc~at the \BiSeEuS\ interface.\cite{Katmis16} 
 \Tb\ of EuS is 16 K and below this temperature the magnetization is in plane. In density functional theory calculations, a magnetization at the \BiSeEuS\ interface was found in the ground state, which may be enhanced by Eu interdiffusion.\cite{Eremeev15} By polarized neutron reflectivity as well as superconducting quantum interference device (SQUID) magnetometry, Katmis et al. \cite{Katmis16} recently suggested that in this system the interface remains ferromagnetic up to temperatures much higher than the bulk \Tc\ of 16 K and persists even to above room temperature. Up to 5 K, a coercivity of about 500 Oe was reported and up to $\sim200$ K it was still observed to be about 50 Oe. Moreover, the ferromagnetic interface was  found to be polarized out of plane while the bulk-like EuS layers are magnetized in the plane below \Tc.\cite{Katmis16} 

Within the exchange coupled divalent Eu monochalcogenide systems with spin-$\frac{7}{2}$, i.e., EuO, EuS, EuSe, and EuTe, the balance between ferromagnetic (FM) and antiferromagnetic (AFM) interactions is known to systematically change from FM in EuO and EuS to AFM in EuTe. 
Topological insulators can show proximity magnetization at interfaces also with antiferromagnets such as in (Bi,Sb)$_2$Te$_3$/MnTe.\cite{He18}
Among the systems closest to EuS, i.e., EuO and EuSe, we have chosen EuSe for our present investigation due to its benign epitaxy behavior as a substrate template, as well as an overlayer for \BiSe.
While transmission electron microscopy data of the \BiSeEuS\ interface reported by Katmis et al. \cite{Katmis16} suggests good epitaxy as well, growth with the same chalcogene anion is much simpler to realize and, in addition, the EuSe lattice constant of 6.196 \AA~\cite{Subhadra92} is by 3.5 \% larger than in the case of EuS. The lattice mismatch with respect to the BaF$_{2}$ substrate (6.20 \AA) used in both works is thus smaller for \BiSeEuSe. Most importantly, the magnetic wide-band-gap semiconductor EuSe displays a complex phase diagram comprising FM, AFM, and ferrimagnetic (FIM) phases due to the delicate balance between the ferromagnetic nearest neighbor exchange and antiferromagnetic next-nearest neighbor exchange.\cite{Lechner05,Sollinger10} Therefore, the magnetic structure of EuSe is expected to be particularly sensitive to small perturbations such as external magnetic fields, exchange bias and strains induced by epitaxial growth.\cite{Lechner05} Thus, it is the perfect material candidate to detect even minute changes in magnetism and exchange couplings induced at interfaces to topological insulators.
 
EuSe crystallizes in the rock salt lattice and its magnetic structure is layered in (111) planes. In zero field, bulk EuSe shows a $\uparrow\uparrow\downarrow\downarrow$ AFM structure which is named type I. The simple AFM structure $\uparrow\downarrow$ of type II also occurs in films with compressive strain.\cite{Lechner05}  In small external fields, both AFM structures undergo magnetic phase transitions into a ferrimagnetic (FIM) phase $\uparrow\uparrow\downarrow\uparrow\uparrow\downarrow$. \cite{Lechner05}  For this reason, our EuSe (111) film template forms a ferromagnetic interface to \BiSe~and is thus, perfectly suited for the investigation of a magnetic proximity effect.
 
The AFM-to-FIM transition temperature of EuSe strongly depends on the external magnetic field and ranges from  2 K at $\sim0.2$ T  to 4.8 K at $\sim0.1$ T.\cite{Lechner05} Interestingly, for an in-plane strain of $-1.1$\%, this transition temperature increases from 4.8 to 6.5 K.\cite{Lechner05}

In the present work, we investigate the magnetic properties of high quality \BiSeEuSe\ heterostructures grown by molecular beam epitaxy. Afer detailed characterization by SQUID magnetometry, we employ interface-sensitive x-ray absorption spectroscopy (XAS) and x-ray magnetic circular dichroism (XMCD) in total electron yield (TEY) to probe the properties of the \BiSeEuSe\ interface. We confirm the divalent configuration of the Eu ions in EuSe and determine the magnetic properties below and above the ferrimagnetic ordering temperature of 5 K. Element-specific XMCD measurements on Eu at 10 K were performed after field-cooling in a true zero-field configuration. 
 
 In all measurements, we do not detect any enhancement of the interface magnetism in this system, in particular no dramatic increase of the ordering temperature, contrary to the enhancement up to room temperature reported for the \BiSeEuS\ system. \cite{Katmis16}

\section{Experimental}

\begin{figure}
	\includegraphics*[scale=0.8]{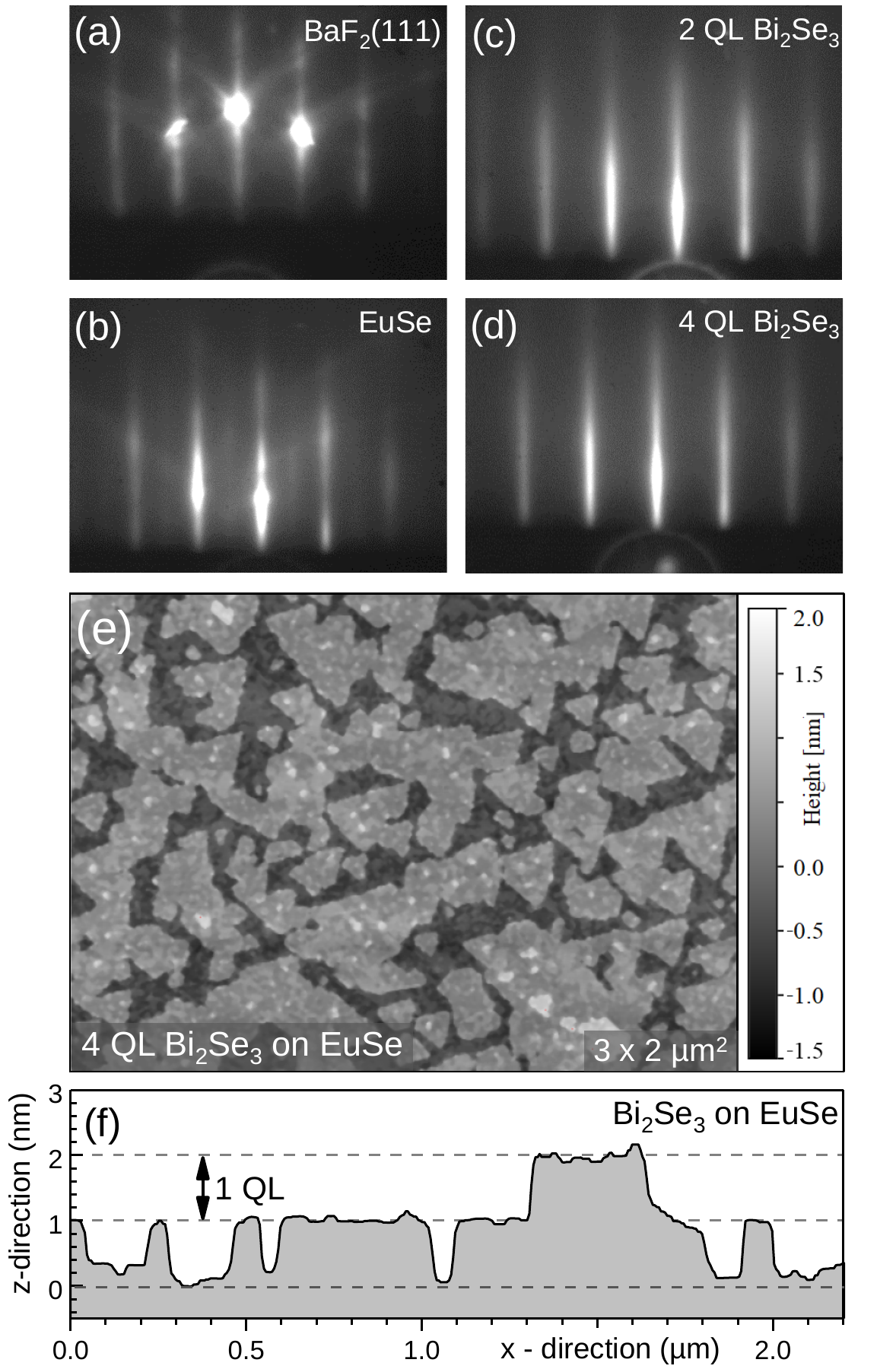}
	\caption{(a-d) Reflection high-energy electron diffraction recorded during the epitaxial growth of \BiSeEuSe~ on the BaF$_2$ substrates at a growth temperature of 260$^\circ$C evidencing a perfect 2D growth mode. (e) Atomic force microscopy image of the sample after deposition of 4 QL \BiSe. The corresponding surface profile is shown in (f) evidencing that a flat surface exhibiting only quintuple layer steps with 1 nm step height.} 
	\label{Fig1}
\end{figure}

Fully relaxed epitaxial EuSe/\BiSe\ films were grown by molecular beam epitaxy on BaF$_{2}$ (111) substrates using \BiSe, Eu and Se sources. EuSe is almost perfectly lattice matched to BaF$_{2}$ \cite{Lechner05} and, thus, EuSe grows expitaxially with the [111] crystal axis normal to the surface. On top of EuSe, \BiSe\ was deposited with a thickness varying between 2 and 8 quintuple layers (QL), forming the \BiSe/EuSe interfaces investigated here. The substrate temperature was 260$^\circ$C and the growth rates 0.2 monolayers (ML)/s for EuSe and 0.067 QL/s for \BiSe\ as determined by quartz microbalance measurements. An excess Se flux of 0.2 ML/s was used in both cases. Under these conditions, perfect 2D growth is achieved as revealed by reflection high-energy electron diffraction (RHEED) shown in Fig.~\ref{Fig1}(a-d), in which perfectly streaked diffraction patterns are sustained at all stages of growth. The final surface was imaged by atomic force microscopy as shown in Fig.~\ref{Fig1}(e), exhibiting atomically flat terraces separated by only by QL steps with 1 nm step height, as seen in the surface line profile shown in  Fig.~\ref{Fig1}(f). 

\begin{figure}
	\includegraphics*[scale=0.8]{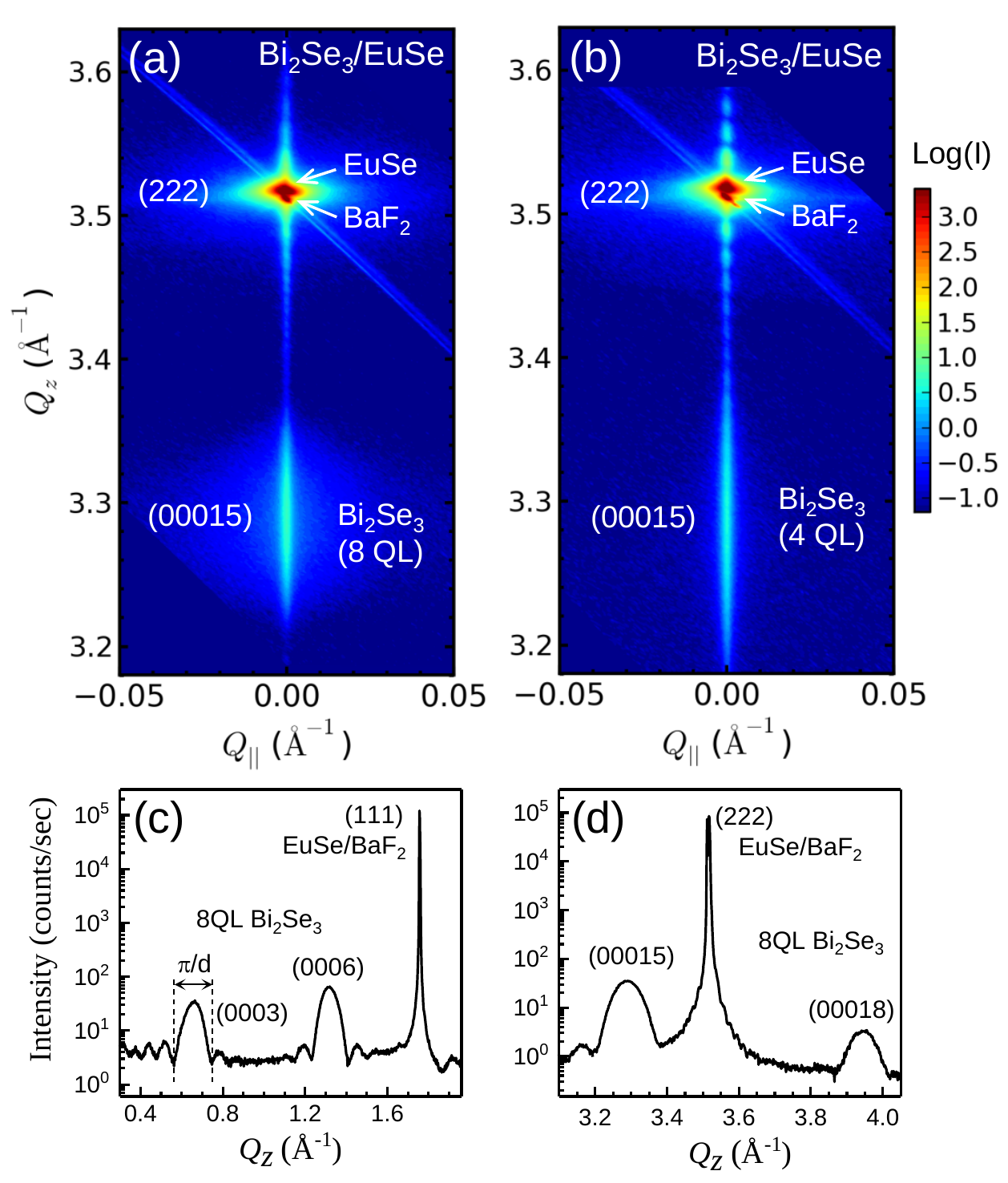}
	\caption{(Color online) X-ray diffraction of 4 and 8 QL epitaxial \BiSe\ on EuSe. (a,b) Reciprocal space maps recorded around the (2 2 2) and (0 0 0 15) reciprocal lattice points evidencing very sharp diffraction peaks for both EuSe and \BiSe\ in the longitudinal (parallel to the surface) $Q_{\parallel}$ direction, where {\bf Q} is the scattering vector with a full width at half maximum comparable to that of the substrate peak. The large broadening along the direction perpendicular to the surface $Q_{z}$ of the \BiSe\ layer peak is due to the very small layer thicknesses of a few nm. (c,d) $Q_{z}$ diffraction profiles of the 8 QL \BiSe\ sample, showing pronouced thickness oscillations around the \BiSe\ peaks from which the layer thickness was derived. The lattice parameters were determined as $a_0$ = 6.188 \AA\ for EuSe (BaF$_2$: $a_0$ = 6.198 \AA ) and $c$ = 28.644 \AA\ for the \BiSe\ layer, both being essentially equal to the bulk lattice constants. } 
	\label{Fig2}
\end{figure}

The high sample quality was verified by high resolution  x-ray diffraction measurements. The results for 8 and 4 QL \BiSe\ on EuSe are depicted in Fig.~\ref{Fig2}, showing in (a,b) the reciprocal space maps around the (222) and (0\ 0\ 0\ 15) reciprocal lattice points of BaF$_2$/EuSe and \BiSe\ are depicted. Evidently, due to the very close lattice matching the EuSe peak almost falls on top of the BaF$_2$ substrate peak and the EuSe peak width is essentially equal to that of the single crystalline substrate. From the peak position the lattice parameter of the EuSe film was derived as $a_0$ = 6.188 \AA\ which is equal to the bulk value.\cite{Lechner05} The \BiSe\ peak appears at lower $Q_z$ value and is extremely narrow in the $Q_{\parallel}$ direction, but due to the very small \BiSe\ thickness is strongly broadened in the $Q_z$ direction. ($Q_z$ is the vertical and $Q_{\parallel}$ the horizontal component of the scattering vector.) Accordingly, pronounced finite-thickness intensity oscillations appear for the \BiSe\ peaks 	as revealed by the $Q_z$ diffraction profiles displayed in Fig.~\ref{Fig2}(c,d). From these oscillations, the \BiSe\  thickness was determined to be in good agreement with the nominal values. The $c$ lattice parameter of the \BiSe\ layer was found to be 28.644 \AA. This is in good agreement with the bulk value, indicating that the \BiSe\ films are completely relaxed. Weak thickness oscillations also appear around the EuSe Bragg  peaks, evidencing the high uniformity and 2D character of all layers in the samples. We did not observe any sample degradation during air exposure and  magnetic measurements, which reveals that the \BiSe\ epilayers effectively protect the EuSe from oxidation in air. In the following, we focus on the magnetic properties of the 4 QL  \BiSe\ on 100 nm EuSe sample, where the \BiSe\ layer is sufficiently thin to access the subsurface EuSe/\BiSe\ interface by XAS and XMCD.

For magnetic characterization, the temperature and field dependences of the magnetization were measured in fields up to 6 T applied parallel and perpendicular to the sample's surface using a SQUID (Quantum Design) vibrating sample magnetometer in DC mode in a wide temperature range between 2 and 300 K installed at the Laboratory for Magnetic Measurements at Helmholtz-Zentrum Berlin. For data collected below 100 K we used a zero-field cooling procedure. SQUID magnetometry of thin film dilute systems requires particular care. This is because already minute amounts of impurity atoms on the sample or holder used in the magnetometry measurements may lead to errorneous conclusions regarding the magnetic state of the material.\cite{Ney11} For these reasons, the recorded magnetic signal was corrected for the substrate and glue contributions that are both diamagnetic.

Moreover, in order to eliminate apparatus effects (coercive field of the superconducting magnetometer coil) and to extract the intrinsic coercitivity of the sample, a pure Pd sample has been measured as a reference, using the same measurement protocol. By this means, the intrinsic coercive field field of our SQUID magnetometer was determined as $\sim$ 30 Oe independent of temperature, which was taken into account in our analysis.

XAS and XMCD spectra were collected using the VEKMAG instrument\cite{Noll17}  in high magnetic fields up to 6 T at the PM2 dipole beamline and the XUV instrument at the UE46-PGM1 undulator beamline of BESSY II. 
The degree of circular polarization is $\sim77$\%\ at PM2 and $\sim90$\%\ at UE46-PGM1 for which the XMCD data have been corrected (except for data shown in Fig.~\ref{Fig10}(a)). 
The experimental data were obtained in the total electron yield (TEY) at both instruments. In the XUV instrument, we field cooled the sample to 10 K in the  stray field of a strong permanent magnet which amounts to about 0.5 T at the sample position. After removal of the magnet, XMCD measurements were performed in zero field.

\section{Results}

\subsection{Bulk magnetometry}

\begin{figure}
\includegraphics*[scale=0.5]{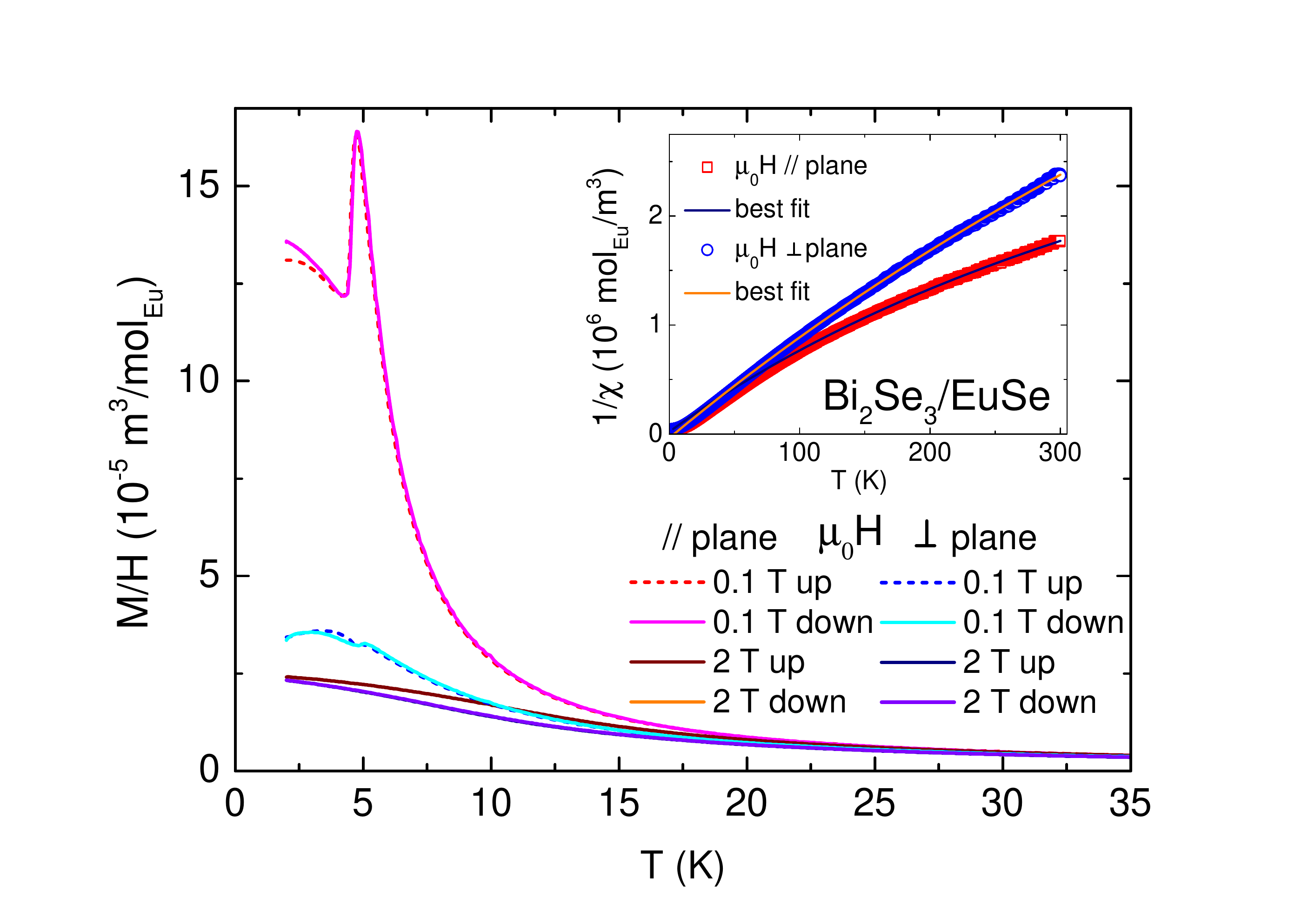}
\caption{(Color online) Low-temperature detail of the temperature dependence of the magnetic susceptibility $\chi (T)$~=~$M(T)$/$H$ measured at 0.1 T and 2 T applied perpendicular and parallel to the surface of the sample with increasing and decreasing temperature. The temperature dependences of the inverse magnetic susceptibility measured in a field of 2 T with increasing temperature for both field orientations are shown in the inset. Full lines through experimental values represent best fits to a modified Curie-Weiss law.} 
\label{Fig3}
\end{figure}

In Fig.~\ref{Fig3} we show the temperature dependences of the magnetic susceptibility $\chi (T)$~=~$M(T)$/$H$, measured at various magnetic fields $H$ applied perpendicular to and along the sample's surface, respectively. As can be seen, the magnetic response is much higher for low fields applied within the plane   than perpendicular to it. At low temperatures, field and history dependent features are visible for both orientations. The magnetic  susceptibility starts to be field dependent below $\sim$ 40 K and for low fields a small differences between zero-field cooled (ZFC) and field cooled (FC) regimes exist. Clear anomalies just below 5 K due to the magnetic phase transitions, visible for both orientations, are in accord with literature data for EuSe.\cite{Lechner05}

At high temperatures above $\sim$ 100 K the magnetic susceptibility follows for both field orientations a modified Curie-Weiss law according to the expression $\chi_{c}$($T$) = $\chi_{0}$ + $C$/($T$ - $\theta_{p}$), where $\chi_{0}$ is the temperature independent term, $C$ denotes the Curie constant and $\theta_{p}$ the paramagnetic Curie temperature. Best fits are shown in the inset of Fig.~\ref{Fig3} by solid lines. The refined values of $\theta_{p}$ = $-2.3$(2) K and 3.7(2) K for the field applied perpendicular and parallel to the surface of the sample, respectively, document that the response of this system to the magnetic field is indeed anisotropic. The effective moment of 8.4(2) $\mu_{B}$ and 8.0(1) $\mu_{B}$ per Eu atom for the field parallel, respectively, perpendicular to the sample surface obtained from the best fits above 100 K are close to the Eu$^{2+}$ free-ion value of 7.94 $\mu_{B}$/Eu$^{2+}$. 

\begin{figure}
\includegraphics*[scale=1.0]{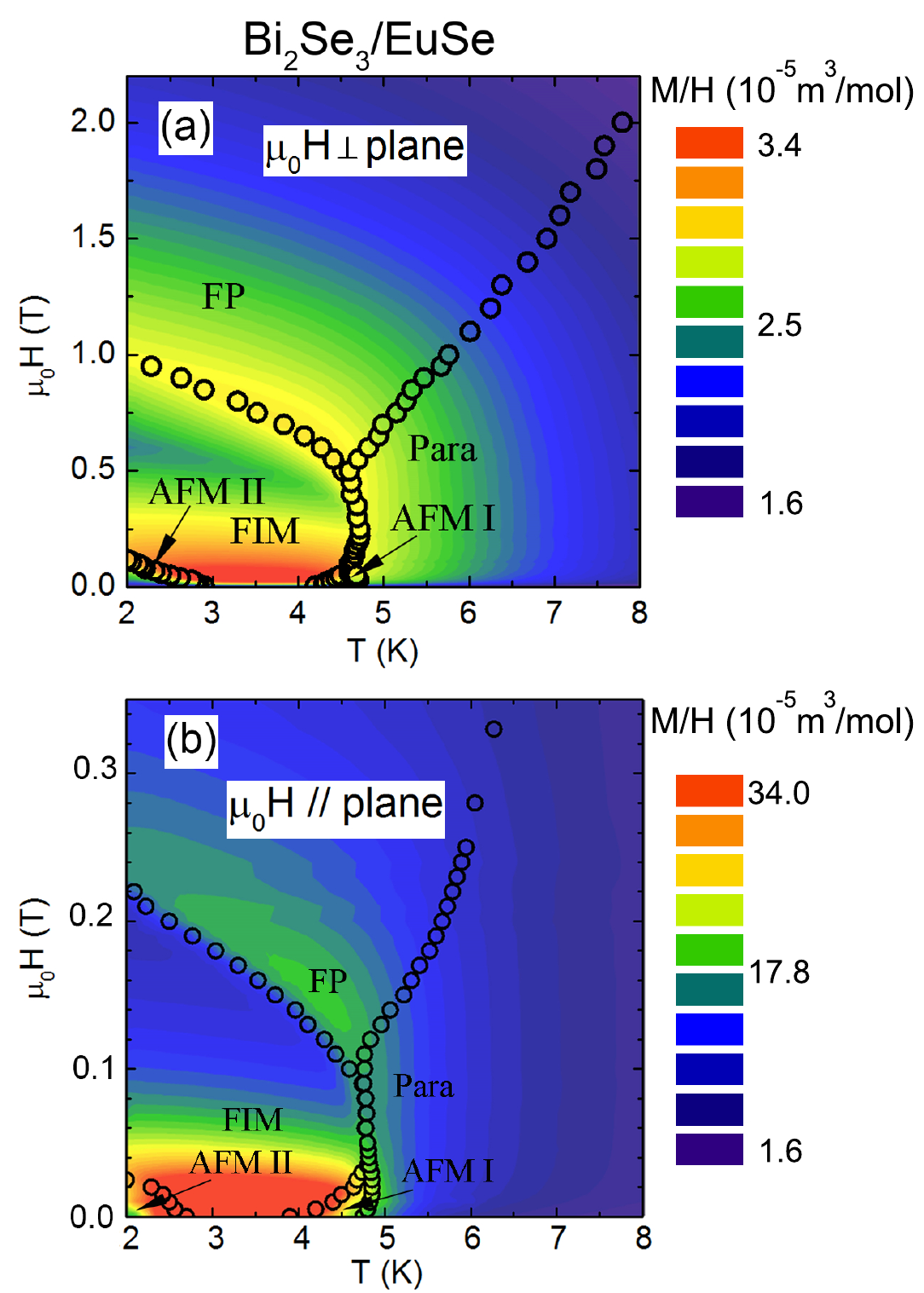}
\caption{(Color online) B-T magnetic phase diagram of \BiSeEuSe~ for field applied perpendicular (a) and parallel (b) to the plane constructed from $M(T)$/$H$ data as described in the main text. The magnetic susceptibility $\chi (T)$~=~$M(T)$/$H$ is color-coded: the blue (red) color denotes low (high) value of $M(T)$/$H$. AFM I, AFM II, FIM, FP, and Para denote the antiferromagnetic ($\uparrow\uparrow\downarrow\downarrow$), antiferromagnetic ($\uparrow\downarrow$), ferrimagnetic ($\uparrow\uparrow\downarrow$), field polarized ($\uparrow\uparrow$) and paramagnetic phases, respectively. Open points denote positions of anomalies marking phase transitions.} 
\label{Fig4}
\end{figure}

In Fig.~\ref{Fig4} (a) and (b) we show the color-coded values of the magnetic susceptibility $\chi (T)$ (portion of the data shown in Fig.~\ref{Fig3}) interpolated to a regular grid from the temperature dependences of the magnetic moment $M(T)$ divided by magnetic field $H$ for the field applied perpendicular and parallel to the (111) sample surface revealing the B-T magnetic phase diagram for the two orientations. Data were taken with SQUID at constant fields with increasing temperature after cooling in zero field. The phase transition lines between AFM and FIM phases and FM and paramagnetic phases manifest themselves as sudden changes in   $\chi (T)$ (in Fig.~\ref{Fig4} visible as color changes). Open points were determined from the temperature derivatives of $T\cdot M(T)$/$H$. The phase diagram for field within the plane is identical to the phase diagrams reported for unstrained EuSe\cite{Lechner05},  proving the absence of any significant strain in our sample.  

At low temperatures below T$_{c1}$ = 2.8(1) K we observe for both field orientations two field-induced phase transitions. These can be observed at 2 K for field perpendicular to the plane at $\sim$ 0.1 T and $\sim$ 1 T and for the field parallel to the plane at $\sim$ 0.02 T and $\sim$ 0.2 T, respectively. The  transition at lower field marks the transition from the type II AFM ($\uparrow\downarrow$) phase towards the FIM ($\uparrow\uparrow\downarrow$) phase. The transition at the upper field is from FIM to a field polarized (FP) phase. For temperatures between 2.8 K and T$_{c2}$ = 3.9(1) K only one transition from FIM to FP is seen for both orientations and for temperatures between T$_{c2}$ and \Tn~= 4.8(1) K, again two transitions are present marking the transition from the type I AFM  phase ($\uparrow\uparrow\downarrow\downarrow$) at low fields to FIM and finally to FP. These findings are in agreement with the literature for EuSe. \cite{Diaz10,Lechner05,Fischer69}

\begin{figure}
\includegraphics*[scale=0.8]{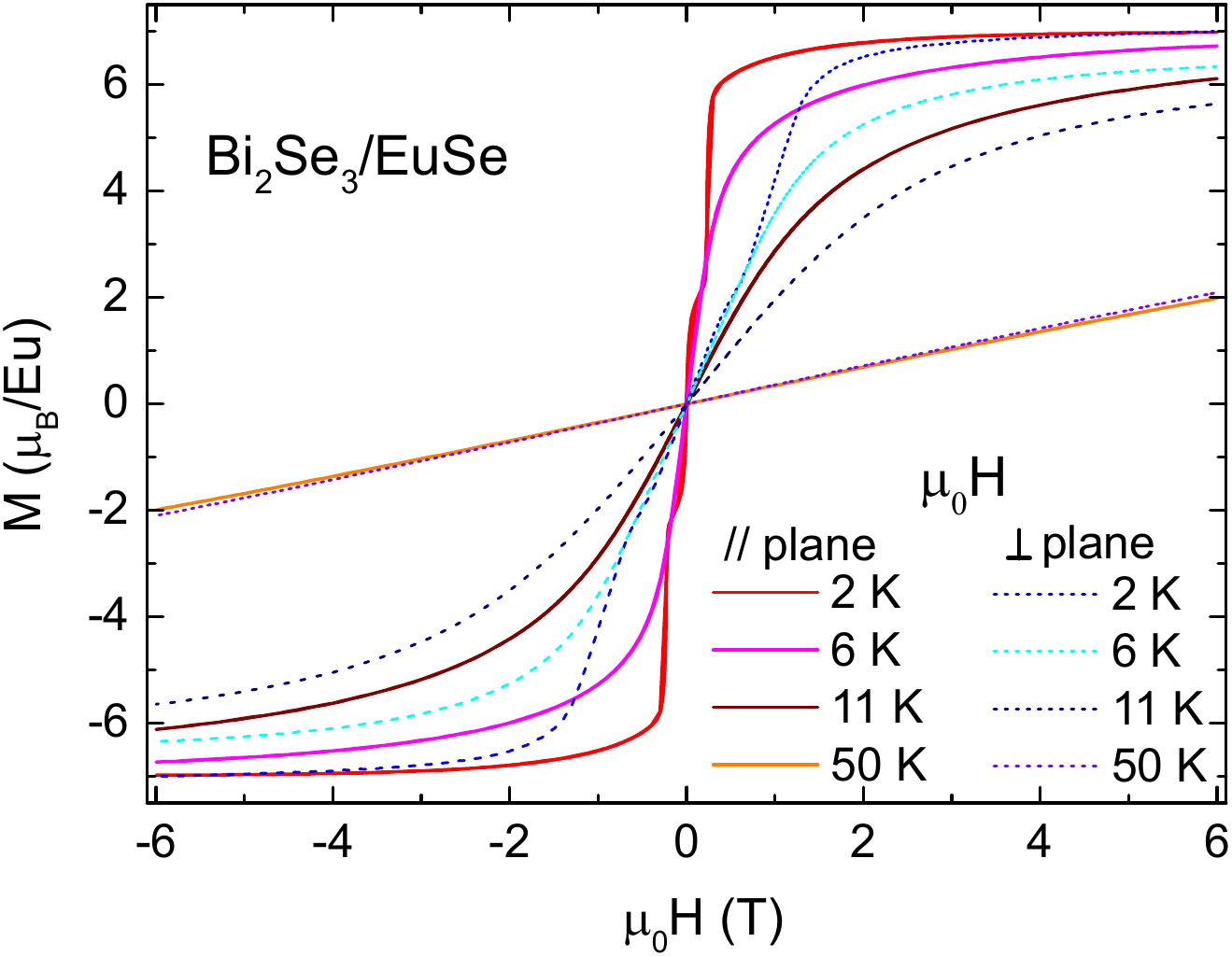}
\caption{(Color online) Field dependences of the magnetic moment per Eu atom measured using SQUID magnetometer at various temperatures between 2 K and 50 K in a wide field range applied perpendicular and parallel to the surface of the sample.} 
\label{Fig5}
\end{figure}

In Fig.~\ref{Fig5} the field dependences of the magnetization for the external field applied perpendicular and parallel to the surface are shown, respectively, documenting that the easy magnetization direction lies within the plane.  For both orientations we observe at low temperatures a clear tendency towards saturation. With increasing temperature, this tendency decreases and above $\sim$ 50 K we observe for both field orientations a linear field dependency of the magnetization. Our SQUID data shows that the net coercive field of the sample obtained from a difference between increasing and decreasing field branches (corrected for the intrinsic behavior of the SQUID superconducting coil) has for field applied perpendicular to the plane above \Tn~between 5 K and $\sim$ 50 K a small non-zero value. This field is below 30 Oe at 10 K and below 10 Oe at 30 K. For the field applied within the plane it is below the detection limit. The remanent magnetization for field perpendicular to the plane amounts at 6 K to about 0.005 $\mu_{B}$/Eu$^{2+}$. This could indicate the existence of an interface magnetization with out-of-plane anisotropy, similar to the one inferred for \BiSeEuS\ .\cite{Katmis16} If true, this effect must be easily detectable by an interface-sensitive method such as  XMCD.

\subsection{Interface sensitive magnetometry}

SQUID magnetometry is a typical bulk probe, averaging over the entire \BiSeEuSe\ heterostructure. To distinguish between interface and bulk magnetism, we therefore employ the more surface/interface sensitive XAS and XMCD in electron yield. In these measurements, the \BiSe\ serves as protection layer that is, on the one hand, thick enough for sample transfer in air without oxidation and, on the other hand, thin enough for detection of the XAS and XMCD signals from the subsurface \BiSeEuSe\ interface in TEY. This is due to the small electron escape depth, for which reason the detected Eu signal stems predominantly from the EuSe layer at the interface. 

\begin{figure}
\includegraphics*[scale=0.5]{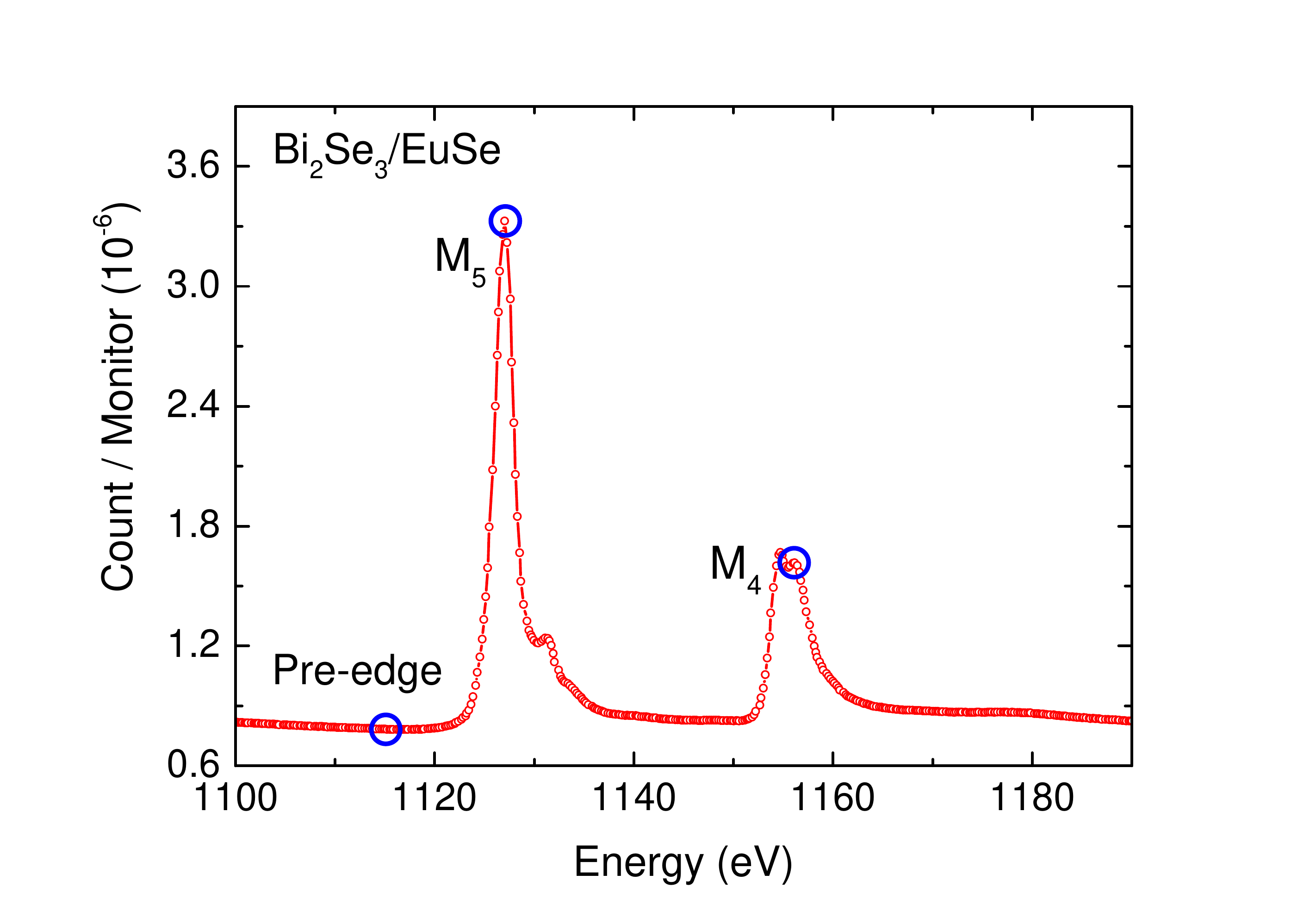}
\caption{(Color online) X-ray absorption spectra of \BiSeEuSe~taken at 2 K using the VEKMAG instrument. The data confirm the Eu$^{2+}$ spin configuration.} 
\label{Fig6}
\end{figure}

In Fig.~\ref{Fig6} we show XAS spectra of \BiSeEuSe~across the Eu M$_{4,5}$ absorption edge taken at 2 K in zero field using the VEKMAG instrument. As can be seen, the shape of the two edges corresponds to the Eu$^{2+}$ valence ground state, proving the good quality of the interface without any Eu$^{3+}$.\cite{Goedkoop88,Selinsky09}

\begin{figure}
\includegraphics*[scale=0.5]{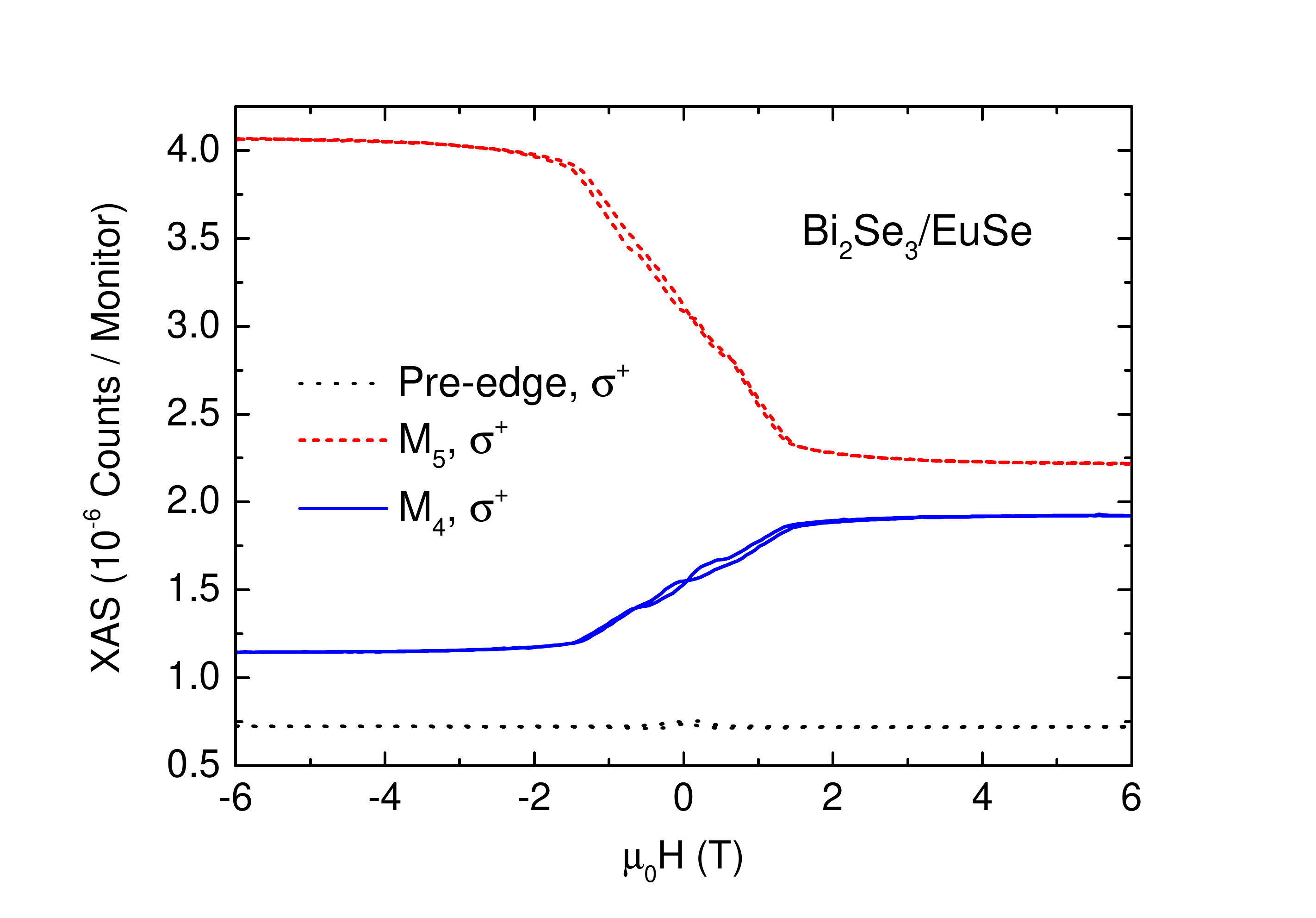}
\caption{(Color online)  X-ray absorption as a function of the applied magnetic field directed perpendicular to the surface of the  \BiSeEuSe~sample taken at 2 K using the VEKMAG instrument at pre-edge (E$_{i}^{PE}$= 1117 eV), M$_{5}$ (E$_{i}^{M5}$= 1129 eV) and M$_{4}$ (E$_{i}^{M4}$= 1158 eV) energies as indicated by circles in Fig.~\ref{Fig6}.} 
\label{Fig7}
\end{figure}

In the next step, we have collected hysteresis curves shown in Fig.~\ref{Fig7}. The incident energies were fixed to values corresponding to pre-edge (E$_{i}^{PE}$= 1117 eV), M$_{5}$ (E$_{i}^{M5}$= 1129 eV) and M$_{4}$ (E$_{i}^{M4}$= 1158 eV),  see Fig.~\ref{Fig6}. The magnetic field has been applied out of plane and changed continuously from 6 T to $-6$ T and then from $-6$ T to 6 T. The data were collected at 2 K using $\sigma^{+}$ and  $\sigma^{-}$ incoming polarizations. While the hysteresis curve taken at the pre-edge energy (dotted line in Fig.~\ref{Fig7}) is almost flat with exception close to  zero field, curves taken at M$_{5}$ and M$_{4}$ edges show a strong field dependence that closely resembles the magnetization curves obtained by SQUID (see  Fig.~\ref{Fig5}). The signal at M$_{5}$ decreases for $\sigma^{+}$  with increasing field, the signal at M$_{4}$ has the opposite field dependence. For the reverted field orientation the the sign of the signal flips. The signal is also reverts for reverted incoming circular polarization $\sigma^{-}$. Such dependences prove that the Eu 4$f$ magnetic moments are aligned by the applied field. Above $\sim$ 1.5 T they show a tendency towards saturation. 

\begin{figure}
\includegraphics*[scale=0.5]{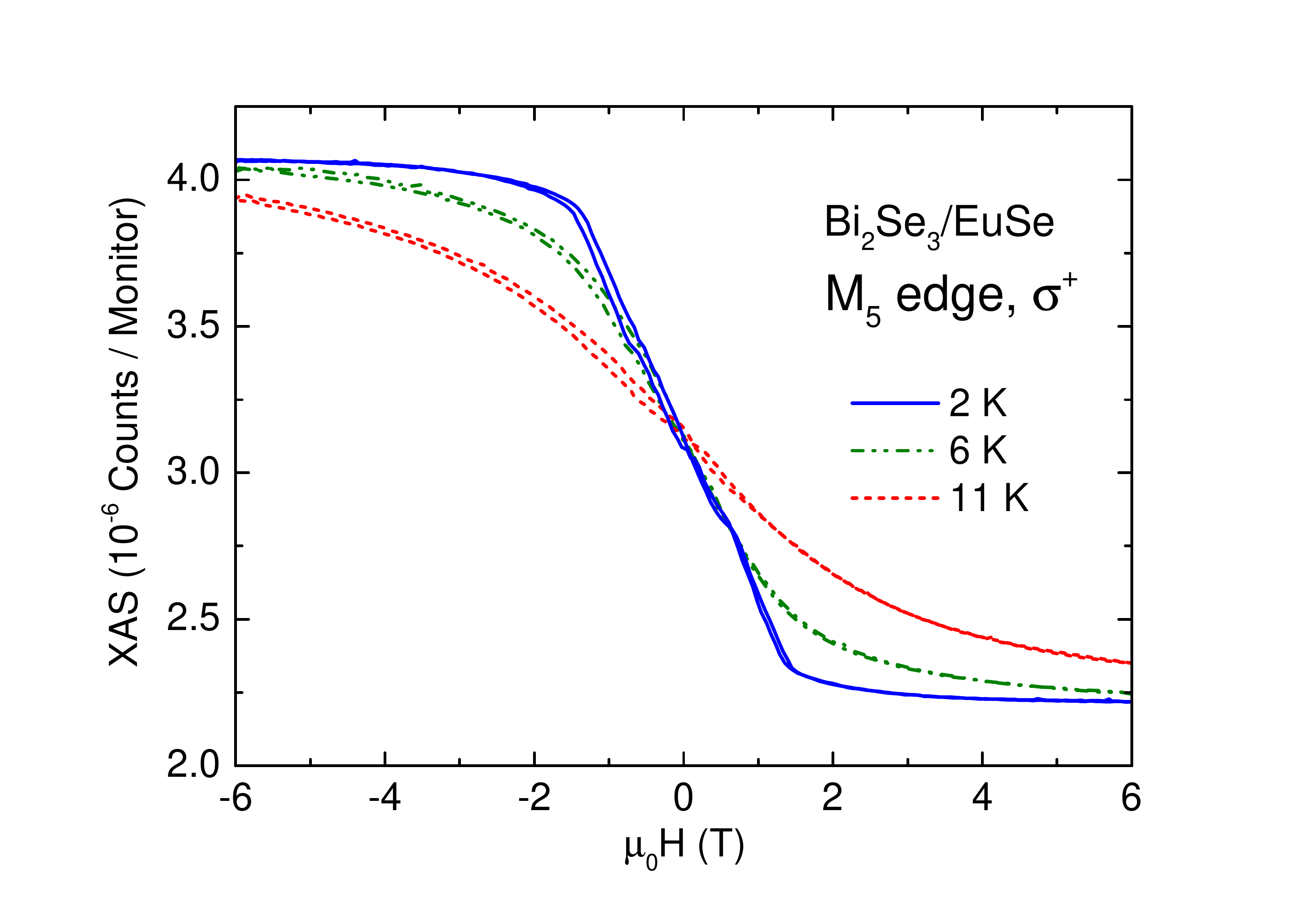}
\caption{(Color online) X-ray absorption as a function of the applied magnetic field directed perpendicular to the surface of the \BiSeEuSe~sample taken at 2 K, 6 K and 11 K using the VEKMAG instrument at the M$_{5}$ (E$_{i}^{M5}$= 1129 eV) edge using incident polarization $\sigma^{+}$.} 
\label{Fig8}
\end{figure}

\begin{figure}
\includegraphics*[scale=0.5]{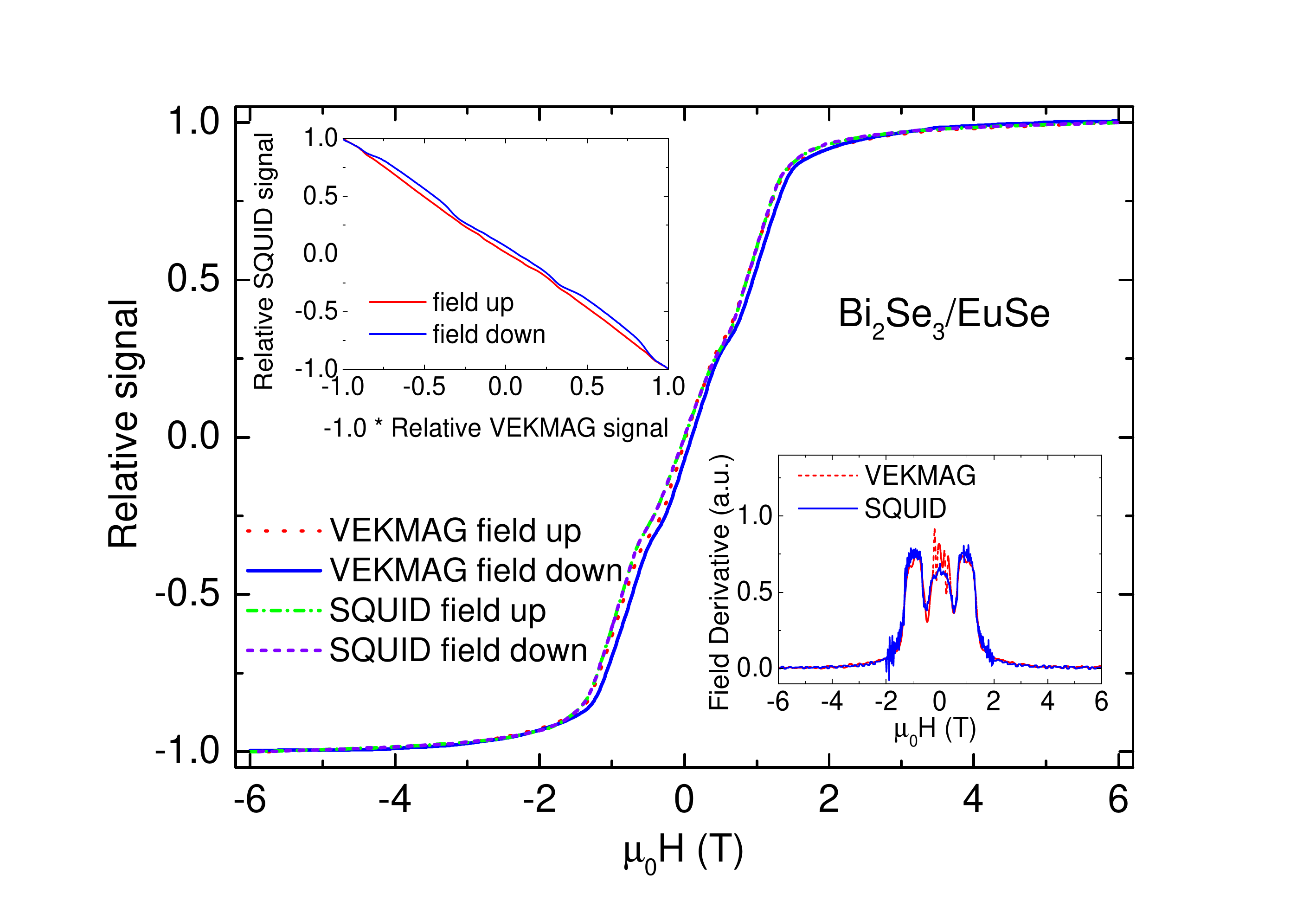}
\caption{(Color online) Comparison between the magnetization curve as a function of the applied magnetic field obtained using SQUID (presented below) and VEKMAG at 2 K. In the left inset we show the direct comparison between SQUID and VEKMAG relative signals. In the right inset we show the field derivative of both signals as a function of applied magnetic field.} 
\label{Fig9}
\end{figure}

In Fig.~\ref{Fig8} we show XAS curves taken at different temperatures of 2 K, 6 K and 11 K at E$_{i}^{M5}$= 1129 eV  using $\sigma^{+}$. While at 2 K we observe sharp changes of the slope around 1.5 T and the same additional small kinks at lower fields as seen by SQUID (Fig.~\ref{Fig5}), these features are not present for measurements at 6 K and 11 K. The good correspondence between the SQUID and XAS data is also documented in Fig. \ref{Fig9}, where we draw the scaled data of both types of measurements on top of each other. As is shown in the lower right inset of Fig. \ref{Fig9}, this applies not only to the relative data but also their derivatives.

\begin{figure}
\includegraphics*[scale=0.6]{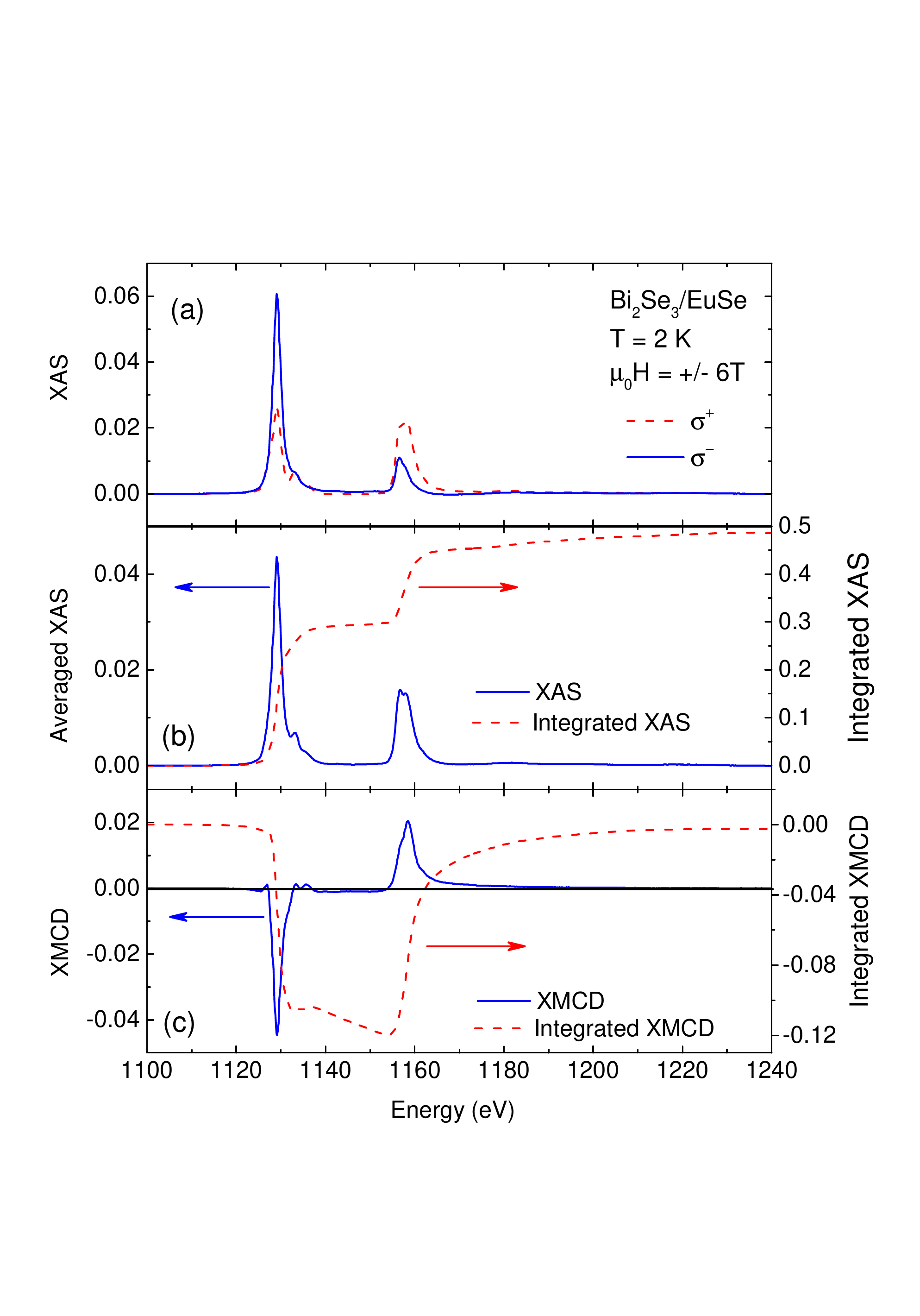}
\caption{(Color online) (a) X-ray absorption spectra of \BiSeEuSe~taken at 2 K using the VEKMAG instrument in an applied field of 6 T measured with opposite circular polarizations. Raw data without correction for the limited light polarization of 77\%\ but after subtraction of a step function (Ref. \cite{Chen95}). (b) XAS spectrum averaged from $\sigma^{+}$ and  $\sigma^{-}$ data shown in panel (a) (left axis) and its energy integration (right axis). (c) XMCD calculated from data shown in panel (a) after correction for the light polarization of 77\%. The data confirm the Eu$^{2+}$ spin configuration.}
\label{Fig10}
\end{figure}

In Fig.~\ref{Fig10} (a) we show XAS spectra of \BiSeEuSe~ taken at 2 K in an applied field of 6 T measured with opposite circular polarizations $\sigma^{+}$ and  $\sigma^{-}$ across the Eu M$_{4,5}$ absorption edge using the VEKMAG instrument. The spectra are not corrected for the incomplete light polarization of 77\%. A step-like background was subtracted using a procedure described in Ref..\cite{Chen95} As can be seen, the shape of both edges corresponds to the Eu$^{2+}$ valence ground state, proving again a good quality of the interface.\cite{Goedkoop88} Fig.~\ref{Fig10} (b) shows the averaged XAS together with its energy integration.  Fig.~\ref{Fig10} (c) represents the XMCD signal derived from the data shown in the panel (a) after correction
for the light polarization of 77\%. In addition, the energy integration is shown. 

Sum-rule analysis \cite{Chen95,Thole92,Teramura96,Carra93} yields spin and orbital Eu magnetic moment of $m_{z,{\rm spin}} = 7.8$ $\mu_{B}$ and  $m_{z,{\rm orbit}} = 0.08$ $\mu_{B}$, respectively. While the orbital part nearly vanishes as expected, the spin part exceeds the atomic  value of the Eu spin-$\frac{7}{2}$ configuration of 7.0 $\mu_{B}$. Considering the systematic uncertainties, the light polarization degree of 77\%\ bears an error of $\pm5$\%. Estimating the error in the integration, we evaluate the parameter $p$ (see Ref. \cite{Chen95}) alternatively near the $M_5$ edge (1135 eV) instead of near the $M_4$ edge (1156 eV) and obtain a reduction of $m_{z,{\rm spin}}$ to $7.4$ $\mu_{B}$. Further sources of systematic error may be a strong magnetic linear dichroism effect \cite{Chen15}
and multiplet effects.\cite{Thole85} We exclude the error in the integration by an alternative analysis method. We analyzed the XMCD asymmetry at the $M_5$ edge and obtain 0.38 before and 0.49 after correction for the light polarization of 77\%. This is by 3\%\ higher than the XMCD in  multiplet calculations\cite{Tripathi18} for Gd$^{3+}$  giving therefore $m_{z,{\rm spin}}=7.2(3)$ $\mu_{B}$. These results confirm that the Eu is indeed in the Eu$^{2+}$ valence ground state. Also at the \BiSe/EuS interface, the Eu spin moment is not expected to change in density functional theory calculations and $m_{z,{\rm spin}}$ = 6.94 $\mu_{B}$ has been predicted\cite{Kim17} in good agreement with our result.

\begin{figure}
\includegraphics*[scale=0.47]{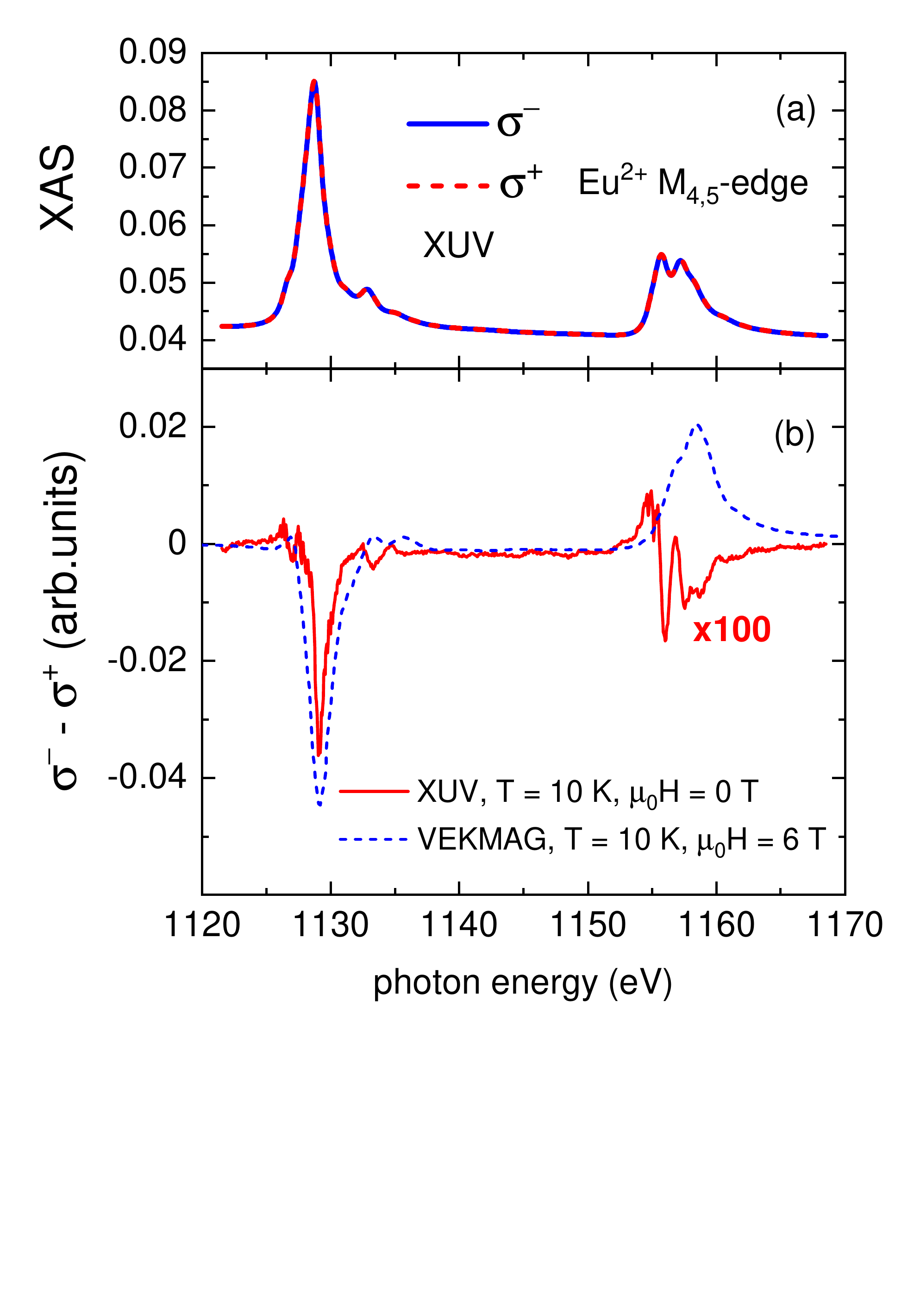}
\caption{(Color online) Comparison of the zero-field data (XUV instrument) to magnetic saturation (VEKMAG instrument) to derive an upper limit for the induced magnetic moment at the interface. (a) X-ray absorption spectra taken with positive $\sigma^{+}$  (red broken line) and negative $\sigma^{-}$ (blue full line) circular polarization at 10 K in zero field using the XUV instrument. (b) Difference between the X-ray absorption spectra taken with positive $\sigma^{+}$ and negative $\sigma^{-}$ circular polarization at 10 K in zero field using the XUV instrument (red solid line) along with the difference between the XAS taken with positive $\sigma^{+}$ and negative $\sigma^{-}$ circular polarization obtained at 10 K under a magnetic field of 6 T using the VEKMAG instrument (blue dashed line, from Fig.~\ref{Fig10}). VEKMAG data have been shifted in energy to meet the energy scale of the XUV instrument due to the known small instrumental differences of the nominal energy scales of different beam lines. Energy scales in panels (a) and (b) are the same. Please, note the multiplication $\times100$ for the zero-field data from the XUV instrument in panel (b).} 
\label{Fig11}
\end{figure}

In Fig.~\ref{Fig11}(a) we show X-ray absorption spectra taken with positive $\sigma^{+}$ and negative $\sigma^{-}$  circular polarization at 10 K in zero field using the XUV instrument after cooling down the sample in the stray field of a permanent magnet. The experimental curves for the two incident polarizations are nearly identical suggesting vanishing Eu magnetism at 10 K in zero field. In Fig.~\ref{Fig11}(b) we compare the XMCD difference recorded at the XUV instrument with the difference XMCD signal obtained at 6 T using the VEKMAG instrument at identical temperature of 10 K. In the case of the XUV data, the sample is in its field-saturated state (see Figs.~\ref{Fig5} - ~\ref{Fig9}). As can be seen, the signal recorded at XUV is negligibly small with respect to the data taken at VEKMAG. These data are shown in Fig.~\ref{Fig11}(b) on a 100 times expanded scale. While these data show small structures at the M$_{5}$ and M$_{4}$ resonance, the line shape is inconsistent with the expectations for a Eu$^{2+}$ XMCD signal but rather reminiscent of an inverted XAS spectrum. We discuss this signal firstly in terms of lineshape and origin and secondly in a quantitative manner. An XMCD difference resembling the XAS lineshape typically occurs when the $\sigma^+$ light and the $\sigma^-$ light cause slightly different intensities due to even tiniest technical inaccuracies. Such artifact is typically corrected by factors but usually the exact normalization procedure can only be known within the experimental uncertainties. In the present case, a factor of  1.004 would lead to a complete disappearance of the XMCD feature at the $M_5$ edge that appears in Fig.~\ref{Fig11}(b). However, it would lead at the same time to a complicated and hardly interpretable lineshape at the $M_4$ edge. In fact, we did not find any normalization that would give a realistic XMCD lineshape with opposite XMCD signs at $M_5$ and $M_4$. From this observation and the comparison between data obtained using VEKMAG and XUV instrument as  shown in Fig.~\ref{Fig11}(b), we can safely rule out the presence of any significant  Eu moment. However, taking the intensity of the M$_{5}$ edge as an upper limit for a potential XMCD contribution to the  measurements with the XUV instrument, we cannot rule out a remanent Eu magnetization of the order of 1/160 of the saturation value of $m_{z}$ = 7 $\mu_{B}$. TEY-XAS is strongly surface sensitive with an exponentially decaying contribution of burried layers with increasing distance from the surface. Typical decay-constants $l_{0}$ are in the order of 2 nm.\cite{Abatte92} However, energy-dependent $l_{0}$ have been reported. \cite{Frazer03} Extrapolating the literature data \cite{Frazer03} to 1150 eV a maximum possible $l_{0}$ = 3 nm can be estimated. Using this value, 97 \% of our observed Eu-XAS stems from the 10 nm extended interface regions to which an enhanced magnetism has been claimed for \BiSeEuS\ \cite{Katmis16} and 12 \% from the topmost Eu atomic layer directly located at the \BiSeEuSe\ interface. Thus, within the quality of our measurements discussed above, we can rule out any average Eu magnetization in the top 10 nm EuSe above 1/160 of the full saturated moment, i.e. $\approx$ 0.04 $\mu_{B}$. Even for the rather extreme situation of a single magnetized Eu layer at the interface, the maximum possible average moment per Eu ion in this layer must be smaller than 1/20 of the saturated moment.

\section{Discussion and Conclusions}

Studying in detail the magnetic properties of a high quality \BiSeEuSe\ heterostructure, we reveal that its magnetic phase diagram perfectly reproduces all features of that of pure bulk EuSe, indicating that the magnetic structure is not influenced by the proximity to the topological insulator epilayer. Focusing in particular on the search for an enhancement of the magnetic response at the interface, analogous to the recent study of \BiSeEuS\, reported to be ferromagnetic even at room temperature \cite{Katmis16}, we do not find such an effect in our system, which due to the nearly compensated AFM and FM exchange interaction, should be particularly sensitive to any perturbations at the interface. In fact, for our \BiSeEuSe\ system we can exclude any coercivities above 30 Oe at 10 K and above by SQUID.

Using XMCD and XAS in electron yield as particularly sensitive probes for the magnetism at the \BiSeEuSe\ interface, we find that the low temperature hysteresis seen by SQUID is well reproduced by the XAS measurements that also confirm the Eu$^{2+}$ configuration in our samples.  We obtain $m_{z,{\rm spin}} = 7.2(3)$ $\mu_{B}$ from XMCD analysis at the M$_5$ edge and $m_{z,{\rm orbit}} = 0.08$ $\mu_{B}$ from a sum-rule analysis. Most importantly, XMCD profiles collected in zero external field after field cooling show the absence of any significant magnetic signal at the Eu$^{2+}$ M$_{5}$ edge at 10 K or above, which reveals that if a magnetic signal is present, it has to be smaller than $\sim$ 0.6 \% of the saturation value, in agreement with the bulk magnetic measurements that probe the integral response of the whole system. Since XMCD is sensitive to the interface only, providing the elemental specific magnetic moment averaged over a small probing depth ($\sim3$ nm), we thus conclude that within our detection limit the magnetization of EuSe is not significantly reduced or enhanced in the proximity to the topological insulator \BiSe . This is highlighted by the fact that due to the nearly balanced FM and AFM exchange interactions, EuSe is particularly sensitive to any perturbations at the surface or interfaces. The difference with respect to the study on \BiSeEuS\ \cite{Katmis16} is the fact that our results are valid for zero magnetic field, whereas experiments on \BiSeEuS\ system have been performed in applied magnetic fields, in particular the polarized neutron reflectometry that was performed at 1 T. 

Note added: Our conclusion of the absence of enhaced magnetism in the \BiSeEuSe\ system is in agreement with the work of Figueroa et al. \cite{Figueroa20} on \BiSeEuS\  that has been published after our submission, in which no such enhancement of magnetism was observed either.

\acknowledgments

We acknowledge Helmholtz-Zentrum Berlin for the allocation of synchrotron radiation beamtime and the use of the Bulk Properties Lab, which is part of the CoreLab Quantum Materials.  
    
\bibliography{Bi2Se3EuSe_bib}

\end{document}